\documentclass[11pt,tightenlines,eqsecnum,floats,aps,nofootinbib,prd]{revtex4-2}
\usepackage{hyperref}
\usepackage{amsmath,amssymb,amsfonts,amsthm,amscd}
\usepackage{enumerate}
\usepackage{colordvi}
\usepackage{units}
\usepackage{epsfig}
\usepackage{natbib}
\usepackage{enumerate}
\usepackage{afterpage}
\usepackage[utf8]{inputenc}
\def\be{\begin{equation}}
\def\ee{\end{equation}}
\def\ba{\begin{eqnarray}}
\def\ea{\end{eqnarray}}
\def\Lie{\mathfrak{L}}
\def\L{\mathcal{L}}
\def\H{\mathcal{H}}
\begin{document}

\title{How closed is cosmology?}

\author{Sean Gryb}
\email{s.b.gryb@rug.nl} 
\affiliation{University College Groningen, Groningen, Netherlands}

\author{David Sloan}
\email{d.sloan@lancaster.ac.uk}
\affiliation{Department of Physics, Lancaster University, Lancaster UK}

\begin{abstract}
\noindent Classical cosmology exhibits a particular kind of scaling symmetry. The dynamics of the invariants of this symmetry forms a system that exhibits many of the features of open systems such as the non-conservation of mechanical energy and the focusing of measures along the dynamical flow. From these properties, we show that important dynamical features emerge that are not present in closed systems. In particular, a large and physically plausible class of cosmological models give rise to a natural arrow of time. We then argue that the appropriate notion of closure in cosmology is dynamical closure - that a system can be integrated without reference to external factors. This is realised in physical systems in terms of the algebraic closure of the equations of motion such that the system is autonomous. Remarkably, in a growing class of models it can be shown that the autonomous system obtained remains regular and can be integrated through the big bang.
\end{abstract}

\maketitle
\section{Introduction}

What is a closed system? In trying to answer this question, we will find standard responses wanting in many respects. Our proposed solution --- one that makes sense for self-gravitating systems --- will have some seemingly unusual features. We will show that these features have unexpected virtues and only seem surprising because of the inadequacies of naive definitions of closed systems. In fact, our proposal will be seen to lead to some rather remarkable conclusions: the universe behaves like an open system where non-conservation of energy leads to the elimination of the conventional Big Bang and a solution to the problem of the Arrow of Time (AoT).

An important aspect of our proposal is the claim that energy need not be conserved in a closed system. Common intuition suggest that if a system is truly closed then there is nowhere for the energy to escape. But such an argument begs the question. Assuming that the energy, which is just some function on the state space of the system, must `escape' somewhere when its value decreases is precisely the assumption that the energy must be conserved.

A more rigorous argument for conservation of energy in a closed system is to require the autonomy of the Lagrangian used to generate the system's laws. Closure of the system could be taken to motivate the formal requirement that the Lagrangian be a function only of an algebraically closed set of dependent variables (e.g., field configurations and their derivatives) with no explicit dependence on the independent variables (e.g., spacetime coordinates $x$ or $t$) of the system. In this case, it's easy to show\footnote{Modulo spatial boundary conditions that could be taken to zero for a closed system.} that the equations of motion are equivalent to a symplectic Hamiltonian system where the Hamiltonian is a function only of phase space. A straightforward application of the theorems of symplectic geometry then suggests that the Hamiltonian function is a constant of motion. Identifying the Hamiltonian with the energy then gives conservation of energy.

This argument, though general and elegant, has several difficulties. First, it's not always appropriate to identify the Hamiltonian function with the energy of the system. Any reparametrization-invariant theory, such as general relativity, will have a vanishing Hamiltonian, and this cannot not lead to an interesting notion of energy. Second, the argument assumes that the laws of the theory are generated by Hamilton's principle, which under the assumption of the autonomy of the Lagrangian leads to a symplectic system. But many interesting physical systems, as we will argue later, are not symplectic. There is thus no guarantee, in general, that the Hamiltonian --- let alone the energy --- will be conserved in a non-symplectic system.

An alternative motivation for conservation of energy could run through Noether's first theorem and the assumption of time-translation invariance of the Lagrangian. Noether's first theorem, however, also assumes that the equations of motion are generated by Hamilton's principle and therefore also excludes non-symplectic systems. Moreover, it's not clear that time-translation invariance of the Lagrangian should be a necessary condition for a closed system. In order to have a definition of a closed system that is general enough to include non-symplectic systems, one must therefore abandon the notion that energy must be conserved.

There are, however, naive arguments against the existence of non-symplectic closed systems. These arguments run through Liouville's theorem, which gives the unique \emph{Liouville measure} that is conserved in a (finite-dimensional) symplectic system.\footnote{No such finite measure exists in a infinite-dimensional system.} The existence of such a measure guarantees that conserved probability distributions can be defined within the theory. Thus, within symplectic systems there is a straightforward way to `conserve probabilities.' The conditions for Liouville's theorem are not, however, necessary conditions for being able to define conserved probabilities outside of symplectic systems. For example, in contact systems, which are not symplectic and will play an important role in our considerations below, generalisations of Liouville's theorem guarantee a unique dynamically conserved measure. Thus, there exists perfectly well-defined notions of probability for closed non-symplectic systems.

Our proposal for defining a closed system does not require the system to be symplectic. Instead, it makes use of an essential element of the proposals given above: the requirement of the autonomy of the dynamical system in question. While symplectic proposals rely on the autonomy of the Lagrangian and the use of Hamilton's principle to generate autonomous equations of motion, ours will focus more directly on the autonomy of the equations of motion themselves. In particular, we will define a closed dynamical system as a system whose dynamical laws are deterministic and algebraically closed in terms of a particular algebra of observables. We will be silent on the nature of the observable algebra in question so that the algebraic closure of our equations does not in principle prevent us from treating probabilistic or quantum theories (modulo the measurement problem).

To begin, we will give a simple example to illustrate different senses in which a dynamical system can be closed. This will help both to motivate our proposal and to illustrate the consequences for cosmology.

\subsection{Closure in the harmonic oscillator}

Let us consider the canonical physical example system of a closed system – the harmonic oscillator, defined by through two constants, the mass $m$ and spring constant $k$, and the equation of motion relating acceleration $a$ to position $x$,
\be a = - k x.\ee
In this system, the mechanical energy, 
\be E = \frac{mv^2}{2} + \frac{kx^2}{2}\ee
which is also numerically equal to the Hamiltonian of the system, is constant. If we plot the trajectory of such a particle on phase space (i.e. plotting momentum against position) we find that the orbits form ellipses, which we can directly see is the result of the energy, $E$, being constant along a classical trajectory. Given an ensemble of such systems, each with some initial position and momentum determined such that the ensemble spreads over some region $R$ in the phase space, the dynamical flow of the ensemble is such that $R$ is rotated about the origin at some fixed angular velocity while the volume it occupies on phase space is constant. This is a direct consequence of Liouville's theorem, which applies to this system.

If our harmonic oscillator interacts with the air surrounding it, we can extend our model to include the effects of air drag. Because of this interaction, we no longer expect to have a closed system. If we assume that the force on the oscillator is proportional to the velocity of the particles, mimicking the effect of air resistance, our equation of motion becomes
\be \label{DHO} ma= - kx - f v\ee 
where $f$ is a constant. The openness of the system is now apparent: The mechanical energy of the system decreases over time because $dE/dt$ is proportional to $-v^2< 0$. The orbits on phase space are no longer ellipses but spirals, which eventually come to rest at the origin. The region $R$ now shrinks over time to the point at the origin.

Although our system is non-conservative and does not preserve measure densities, it is autonomous because the equations of motion are entirely self-contained: Given an initial $x$ and $v$ and the value of $f$, we can solve the dynamical equations for all time without reference to any other information. Had we wished to include the effect of a more complicated interaction with air, such as wind blowing through the system, we would no longer be able to describe the system just in terms of its intrinsic properties, but would rather need extrinsic information such as the speed and direction of the wind, and thus our system would lose its autonomy. However, if our damped oscillator is to be taken as a complete system making no reference to the outside world, the value of $f$ would be akin to a constant of nature. Given access to measuring equipment, such as an idealized rod and clock, the value of $f$ could be determined by observing the motion of the oscillator. The system can thus be seen to be autonomous in that all future behaviour can be inferred from current measurements. 

Importantly, the models treated in this section illustrate three distinct notions of `closure' of a particular dynamical system. It can be energetically closed, in which case energy is conserved. It can have a preserved notion of typicality, in which case the phase-space measure density is conserved. And finally, it can be dynamically isolated, in which case the evolution is autonomous. Note that these notions are completely independent: physical systems may exhibit any or none of these. But because the implications of each notion can differ drastically, it is important identify which notions are relevant (or not) to the system in question.

\section{Contact Systems}

One class of systems with autonomous equations of motion that gives rise neither to conserved Liouville forms nor to conserved energies is the class of contact systems. Contact mechanics can be thought of as the odd-dimensional counterpart to symplectic mechanics. Here, for reasons of length, we will introduce only the necessary parts for our analysis, and work directly in Darboux coordinates. Comprehensive and general introductions can be found in \cite{Bravetti,sean_thesis}. 

Herglotz \cite{Lazo_2018} considered a generalization of the usual Lagrangian mechanics to include Lagrangians that are dependent on the action, i.e.
\be S = \int \L(q_i,\dot{q}_i,S) dt \ee
the minimization of which gives rise to a slightly modified version of the usual Euler-Lagrange equations, henceforth called the Herglotz-Lagrange equations:
\be \frac{d}{dt} \left(\frac{\partial \L}{\partial \dot{q}_i}\right) - \frac{\partial L}{\partial q_i} - \frac{\partial \L}{\partial S} \frac{\partial \L}{\partial \dot{q}_i} \ee
Here we see that when the Lagrangian is independent of the action the usual equations are recovered. The Lagrangian is taken to be a function of the tangent bundle over configuration space and the action. This is an odd dimensional space. By construction, our system is autonomous because the Lagrangian is a function only of the coordinates, their derivatives and the action. We can perform a Legendre transform and arrive at a contact Hamiltonian following the usual scheme. Let $p = \frac{\partial \L}{\partial \dot{q}}$. Then, the contact Hamiltonian is
\be \H^c = \sum_{i=1}^n p_i \dot{q}_i - \L \ee
From this have the contact Hamilton equation:
\be \dot{q}_i = \frac{\partial \H^c}{\partial p_i} \quad \dot{p}_i = - \frac{\partial \H^c}{\partial q_i} - \frac{\partial \H^c}{\partial S} \frac{\partial H^c}{\partial p_i} \quad \dot{S} = \sum p_i \frac{\partial \H^c}{\partial p_i} - \H^c \ee
Again, the usual Hamilton equations are recovered when the contact Hamiltonian is independent of the action. It is a straightforward exercise to see that the contact Hamiltonian is not conserved under time evolution with 
\be \dot{\H}^c = -\frac{\partial \H^c}{\partial S} \H^c \ee
The contact form, $\eta=\sum_{i=1}^n p_i dq_i-dS$, is the odd-dimensional counterpart to the symplectic potential. To obtain a volumr form on our contact phase space, written in the coordinates $q_i$,$p_i$ and $S$, we take 
\be \label{contactmeasureeom} \Omega^c = \eta \wedge d\eta^n = -dS \wedge \bigwedge_{i=1}^n (dq_i \wedge dp_i) \ee
Since the space is odd dimensional, we cannot construct a Liouville-type measure because the symplectic form is a two-form, and hence any product of symplectic forms is an even order form that cannot serve as a volume form for an odd-dimensional space. In contrast to the Liouville form, the privileged volume form on a contact manifold is \emph{not} conserved under time evolution and obeys 
\be \dot{\eta} = -\frac{\partial \H^c}{\partial S} \eta \rightarrow \dot{\Omega}^c = -(n+1)\frac{\partial \H^c}{\partial S} \Omega^c \ee
Let us illustrate our construction by considering the damped harmonic oscillator. If we begin with the Lagrangian
\be \L = \frac{m\dot{x}^2}{2} - \frac{kx^2}{2} - \frac{\mu S}{m} \ee
then the Herglotz--Lagrange equation for $x$ is precisely that of the damped harmonic oscillator \ref{DHO}. Identifying $p=m\dot{x}$, the contact Hamiltonian is thus
\be \H^c = \frac{p^2}{2m} + \frac{kx^2}{2} + \frac{\mu S}{m} \ee
It is important to note that this does not correspond to what we would think of as the mechanical energy of the oscillator as it contains a term proportional to the action. The evolution of the contact Hamiltonian shows that it has an exponential decay, as does the contact volume-form.

We can interpret the decay of the contact volume-form as a focusing of solutions on the contact space. This focusing of solutions, in turn, introduces a local time asymmetry into the system along particular solutions, and thus can be used to define a local Arrow of Time. More globally, because this focusing is monotonic along \emph{all} solutions and points in the same direction as the decrease in the contact Hamiltonian, the asymptotic future of any damped oscillator is a system at rest, $x=0=\dot{x}$. The future of all solutions converges on to this point. Hence, for all solutions we can further identify a global Arrow of Time: the direction pointing towards this final state. Equivalently, this Arrow of Time is the direction in which the energy (either mechanical, or the contact Hamiltonian itself) is decreasing. See \cite{sean_thesis} for an extensive treatment of how an arrow of time can emerge through such a mechanism. The damped oscillator thus has a qualitative feature distinct from its undamped counterpart. 

A given contact system can be symplectified. Doing so will produce a symplectic system in which the Hamiltonian is conserved under evolution. This is achieved at the cost of embedding the contact phase space into a phase space with one extra dimension. This process will imply that a single solution to the contact system is mapped onto multiple solutions of the symplectic system, which will appear as a symmetry in the symplectic representation. A simple way to achieve this is to expand our configuration space by introducing a new coordinate $Q$ orthogonal to the contact space. We can then define the symplectic Hamiltonian and symplectic potential by 
\be \H^s = Q \H^c \quad \theta = Q \eta \ee
It is a simple exercise to show that both the symplectic Hamiltonian and potential are conserved as a consequence of Hamilton's equations, and that the equation of motion for each of the contact variables remains unchanged by this process. In our damped harmonic oscillator example, our symplectic system is then
\be \H^s = \frac{\pi^2}{2 m Q} + \frac{k Q x^2}{2} + \frac{\mu Q S}{m} \quad \omega = dS \wedge dQ + d\pi \wedge dx  \ee
Note that the observable algebra of this system is not generated by the original $x$ and $p=v$ variables but $x$ and $\pi= Qv$. This system, as any system made from such symplectifications, exhibits dynamical similarity of order 1 under the similarity vector $D=Q \frac{\partial}{\partial Q} + \pi \frac{\partial}{\partial \pi}$, i.e 
\be \Lie_D \H^s = \H^s \quad \Lie_D \omega = \omega\,. \ee
This reflects the redundancy brought to the system through the introduction of the coordinate $Q$. 

At this point it is natural to ask why, if an open (contact) system can always be embedded within a closed (symplectic) system of this type, do we not simply declare the closed system to give a complete description? Our answer comes from demanding autonomy only on the smallest algebraic set required to retain empirical adequacy \cite{Gryb:2021qix}. Let us motivate this demand now.

In our symplectification process we have, by necessity, introduced an auxiliary structure $Q$ into the system. In the case of the damped harmonic oscillator, the momenta depend upon this auxiliary variable $Q$. Similarly, both the conserved volume-form and the conserved energy (ie, the symplectic Hamiltonian) are also dependent upon $Q$, which can take arbitrary values. All these quantities will therefore take arbitrary values themselves, and thus cannot be used to represent physical structures. The value of $Q$ can therefore have no empirical significance. We see that autonomy in the symplectic system is only achieved in terms of an algebra that is strictly larger than the minimum required for empirical adequacy. In terms of the measure, this means that states are counted as distinct even if they differ only by the (arbitrary) value of $Q$. This is clearly not a desirable state of affairs as it would introduce a distinctions without a difference.

In the case of the damped oscillator, we thus find that the non-conservative, contact formulation should be treated as a complete description of the system as it involves an autonomous system that is independent of auxiliary structures. We conclude that certain apparently `open' systems can reasonably be considered to be `closed' in the sense that they provide complete descriptions.

\section{Cosmology}

Let us now apply our framework to cosmology. In our analysis we shall restrict to the simplest case: flat ($k=0$) Friedmann-Lema\^itre-Robertson-Walker (FLRW) models. The results can be generalized to include open ($k=-1$) and closed ($k=1$) spatial slices, and include anisotropies, together with general matter sources \cite{Sloan:2020taf,Mercati:2019cbn,Sloan:2022exs}. There is an unfortunate clash of nomenclature here wherein cosmologically the terms `open' and `closed' (and separately `flat') are used to refer to the geometry of a spatial manifold, which can either be a three dimensional hyperboloid, a three-sphere or a compactified section of $\mathbb{R}^3$. The issue of whether or not a spatial slice is compact is not at issue here since the stringent demands of homogeneity and isotropy render such considerations orthogonal to the issue of `open' versus `closed' in terms of conservation laws or autonomy.

Cosmology provides an interesting test case for distinguishing between the various demands of being `closed'. Since the object under consideration is taken to be the entire universe, it may be argued if some quantity is to be conserved in a complete accounting of physics, then there should be nowhere to hide any losses. However, one may counter that cosmology is simply the consideration of long wavelength modes of a full theory of general relativity, and that a complete model should account for the transfer of, e.g., energy between modes of differing wavelengths. Thus, a failing of strict conservation laws within the cosmological context does not rule out such conservation for the full theory. Such an arguments does not appear to hold in the vicinity of space-time singularities where, per the Belinski-Khalatnikov-Lifschitz \cite{BKL,Berger,Ringstrom,Uggla,Uggla2} conjecture, the dynamics of the full theory of general relativity asymptotes to those of a homogeneous solution at each spatial point. Further, since cosmological solutions are exact solutions of the full theory, as opposed to truncations in which some terms have been disregarded, any features of these solutions are necessarily features of at least a subset of solutions to the full theory. 

For our purposes it will suffice to demonstrate our results in the flat FLRW model with simple scalar fields as matter. The behaviour of such a cosmology can be found from the Einstein--Hilbert action following a symmetry reduction to homogeneous and isotropic subsystems. After some technical considerations, the foremost of which is checking compatibility with the principle of symmetric criticality, we can render the action as 
\be \label{EHAction} S_{\rm EH} = \int v \left(-\frac{\dot{v}^2}{3v^2}  + \frac{\dot{\phi_i}^2}{2} - V \right) dt\,, \ee
where $V=V(\vec{\phi})$ is a potential function and, to streamline our algebra, we adopt the convention $8\pi G=1$. We here note that a cosmological constant can be included trivially in this system by having a non-zero minimum of $V$. The geometric sector is described by $v$, which represents the volume of a fiducial cell in comoving coordinates. Note that although the scale factor is often referred to as being the `size of the universe', this is undefined in the case of an infinite spatial slice. 
From this action we can derive (following a Legendre transformation) the Hamiltonian description of our theory. We write this in terms of a Hamiltonian function and symplectic structure on phase space
\label {SympHam} \ba 0:=\H &=& v\left(-\frac{3P_v^2}{8} + \frac{P_i^2}{2v^2} + V \right) \\
\nonumber \omega &=& dP_v \wedge dv + dP_i \wedge d\phi_i\,. \ea
Here, $P_v = -\frac{4\dot{v}}{3v}=3H$, where $H$ is the Hubble parameter. Note that the Hamiltonian is a constraint and must be zero --- a fact that follows from the time reparametrization invariance of general relativity. From the existence of these structures it is easy to verify that the `energy' (the Hamiltonian) and the measure $\Omega=\omega^{1+n}$ where $n$ is the number of scalar fields, are conserved over time. Furthermore, from Hamilton's equations we can find an equations of motion of each of the phase-space variables written in terms of itself and the other phase-space variables. Our system is therefore autonomous. It would hence seem apparent that cosmology is a closed system following all three notions that we have discussed.

There is, however, a catch: The algebra generated by the set of variables just described is strictly \emph{larger} than what is needed for empirical adequacy. In particular, the volume $v$ can be set to any value by re-scaling its initial condition without affecting any empirical predictions of the theory. This is because such predictions only ever rely on the value of $v$ relative to some reference value (usually taken to be an initial or final condition).

The equations of motion of our system are the Friedmann equations and the Klein-Gordon equation:
\ba H^2 &=& \frac{1}{3} \left(\frac{\dot{\phi}_i^2}{2}+V \right) \\
    \dot{H} &+& H^2= - \frac{1}{3} \left( \dot{\phi}_i^2 + V \right) \\
    \ddot{\phi}_i &+& 3H \dot{\phi}_i +\frac{\partial V}{\partial \phi_i} = 0 \ea
It is clear from these that the dynamical algebra closes in terms of the $\phi$ and $H$ without ever needing to refer to $v$. We should therefore seek a description of our system that never makes reference to $v$ in the first place. To do so, we note that the freedom to choose the value of $v$ at any time represents a scaling symmetry, and thus we can follow the analysis of \cite{DynSim,Bravetti_2023,sean_thesis} and reduce our symplectic system to a contact system with one fewer dimension. In fact, it was shown in \cite{Sloan:2020taf,Sloan:2022exs} that the dynamics of this system can be derived from an action principle that never refers to scale in the first place by extremizing the action 
\be S_h = \int_0^t \left( \frac{3S_h^2 (t')}{2} + \frac{\dot{\phi}_i^2}{2} - V \right) dt'\,. \ee
This is an action of the Herglotz type as discussed above. Together with the identification $S_h=-H$ this produces the original equations of motion. The action is equivalent to the second Friedmann equation and the Herglotz--Lagrange equation for $\phi$ is the Klein--Gordon equation. The contact Hamiltonian is then equivalent to the first Friedmann equation:
\be 0:=\H^c = -\frac{3S^2}{2} + \frac{p_i^2}{2} + V \quad \eta = -dS + dp_i \wedge d\phi_i \ee
with $p_i=\dot{\phi}_i$. 

There are several notable features of this description of our system. The contact Hamiltonian and contact structure are expressed directly in terms of the minimum autonomous system ($\phi, \dot{\phi}, S=-H$) required for empirical adequacy rather than the usual symplectic variables used in cosmology. The identification of the action $S$ with the Hubble parameter makes the nature of `Hubble friction' more apparent. From equation \ref{contactmeasureeom}, we see that the natural volume form over physical quantities will shrink in time. This solution focusing is a hallmark of `open' systems.

The action is monotonically increasing, and can therefore act as a physically observable proxy for time. It is common when addressing issues of measure in cosmology to construct a space of solutions by restricting to a surface whereon the Hubble parameter is constant \cite{Gibbons:1986xk,Measure}. This is the restriction to a constant extrinsic curvature slicing in the cosmological context. Doing so demonstrates clearly the distinction between the symplectic and contact cases: If we restrict the Liouville form in each case to this slice, we find that $\Omega^s = \dot{\phi} dv \wedge d\phi$ whereas $\Omega^c =\dot{\phi} d\phi$ since the contact phase space is one dimension smaller. Thus in the symplectic case, when evaluating measures on phase space, one must choose an interval in $v$ over which to integrate. This can lead to problems because the volume is not empirically significant. However, since the amount by which volume changes between two times is, the freedom to chose different initial conditions for $v$ leads to ambiguous statements \cite{Measure2,Sloan:2015bha,Corichi:2010zp,Corichi:2013kua}. In the contact case, the measure is unambiguous but not preserved under time evolution. The time evolution of the measure is proportional to the action, and hence solutions focuses on those which maximize the integral of the Hubble parameter between two time slices. 

These considerations show that in cosmology, as in the damped oscillator, `open' dynamical systems can reasonably be understood to give complete, though idealised, descriptions of the universe.

A further enticing feature of the open description of cosmology is that the system of differential equations it produces remain deterministic at, and beyond, the initial big bang singularity. The status of singularities within general relativity as a whole is a complex topic with the Hawking-Penrose theorems proving that, under fairly generic circumstances, there will exist points in a space-time beyond which the paths of physical objects within the system become non-deterministic as the geodesics that objects follow are incomplete. Thus, Einstein's equations do not admit a unique continuation beyond such points. In the cosmological setup, our considerations are somewhat simpler: we need only concern ourselves with the determinism of the system of equations derived from the symplectic Hamiltonian description, equations \ref{SympHam}. The relevant point in evolution is when the Hubble parameter becomes infinite as does (necessarily) at least one contribution to the matter energy density. A theorem due to Foster \cite{Foster} shows that under a broad range of conditions this corresponds to the kinetic energy of the scalar field becoming infinite. 

Infinities, even of physically observable quantities, do not necessary signal the breakdown of the evolution equations. If we consider the harmonic oscillator described in terms of $y=1/x$, we see that its evolution results regularly in infinities of $y$ in finite time. The physical system, with $x$ interpreted as the distance to equilibrium, sees such points as nothing special and continues beyond them. The infinities may appear, on first glance, to inhibit the integration of the equations of motion. However, when rendered in the usual manner through a change of variables, we see that they are in fact an artefact of the description.

In our cosmological model it is the scalar fields that become infinite. We can change coordinates through a gnomonic projection. For two scalar fields, this is achieved by projecting a point on a plane onto one half of a sphere. The unit radius sphere is tangential to the plane, touching at the origin. Each point $P$ on the plane is then mapped uniquely to a point $Q$ on the lower half of the sphere by considering a line from the center of the sphere to the point $P$, and taking $Q$ to be the point of intersection of this line with the sphere. Thus two scalar field values, $\phi_1$ and $\phi_2$ are represented as a pair of longitude and latitude on the sphere. The singularity in the original coordinates (where at least one of the fields has become infinite) is mapped to the equator of the sphere. It is then a matter of technical calculations \cite{Through,Sloan:2019wrz} to show that the equations of motion of the system remain well defined and predictive when expressed in terms of these sphere based coordinates.

\section{Conclusions}

Throughout this work, we have investigated different inequivalent ways in which dynamical systems can be considered closed. These different notions include closure in the sense of conserved energy, conserved measure density and dynamical autonomy. While we found merits for each, no single notion of closure was completely adequate for all purposes. In particular, we found conventional notions of conserved energy and measure to be wanting when applied to the cosmological setting. In that context, we compared a closed (in the sense of preserved measure density and Hamiltonian) symplectic description of cosmology to an open contact formulation. We found that the latter should be taken to give a better complete description of the system.

Central to our argument was a requirement to treat as physical only those structures that are strictly necessary for maintaining empirical adequacy. The symplectic `closed' descriptions violate this principle because they treat as distinct states that differ only by the overall scale of the universe (as determined by $v$), which is not empirically relevant. In contrast, the contact description does satisfy this principle as it excludes $v$. While this description does not conserve energy or measure density, it is nevertheless autonomous in the empirically sufficient variables, and is thus `closed' in the third sense discussed above. This suggests that the appropriate notion of closure in the cosmological setting is that of dynamical autonomy.

In addition to this epistemic argument favouring the contact description, we also showed that the contact description of cosmology has several other explanatory advantages over the standard symplectic description. Firstly, the contact description provides a natural dynamical explanation for the Arrow of Time. This arrow comes from the focusing of solutions along the dynamical evolution --- particularly when this focusing is monotonic across a dense set of solutions that are consistent with observation. Indeed, we saw that this is precisely what occurs in the cosmological models we considered since the solution focusing is proportional to the Hubble parameter, and Hubble expansion is monotonic under a dense set of solutions in the presence of a positive cosmological constant.\footnote{ Particularly when the matter content is required to obey any of the energy conditions of general relativity. } In symplectic systems of cosmology, no such explanation is available so that the Arrow of Time is usually explained in terms of some version of a past hypothesis. The limitations of a past hypothesis have been discussed by many authors (e.g., \cite{earman2006past,gryb2021new}). In the description we are advocating, these difficulties are avoided because a past hypothesis is unnecessary.

A second advantage of the `open' model\footnote{ Here, we mean `open' in the sense of having no conserved measure density and Hamiltonian but closed in the sense of dynamical autonomy. } is that, while it agrees with the `closed' description wherever that description is valid, it continues to be regular when the `closed' system breaks down. The `open' model can be integrated through the initial singularity unambiguously. The contact system is thus sufficient to save the phenomena while avoiding the problematic aspects of a big bang singularity.

\bibliographystyle{apa-good}
\bibliography{OpenCosmo}{}

\end{document}